\documentclass[aps,pra,amsmath,reprint,floatfix,superscriptaddress]{revtex4-1}
\usepackage{microtype} 
\usepackage{amssymb}
\usepackage{graphicx}
\usepackage{enumerate}
\usepackage{bm}

\RequirePackage[
  hyperindex,colorlinks,bookmarksnumbered,
  plainpages=true,pdfstartview=FitH]{hyperref}
\hypersetup{linkcolor=blue,urlcolor=blue,citecolor=blue} 
\usepackage{hyperref}
\usepackage[all]{hypcap}

\newcommand{\ED}{.}
\newcommand{\EC}{,}

\renewcommand{\sp}{^\textrm{sp}}
\newcommand{\orb}{^\textrm{orb}}
\newcommand{\ch}{^\textrm{ch}}
\renewcommand{\vec}[1]{\bm{#1}}

\begin{document} 

\title{Orbital differentiation in Hund metals}
\author{Fabian B.~Kugler}
\affiliation{Arnold Sommerfeld Center for Theoretical Physics, 
Center for NanoScience,\looseness=-1\,  and 
Munich Center for \\ Quantum Science and Technology,\looseness=-2\, 
Ludwig-Maximilians-Universit\"at M\"unchen, 80333 Munich, Germany}
\author{Seung-Sup B.~Lee}
\affiliation{Arnold Sommerfeld Center for Theoretical Physics, 
Center for NanoScience,\looseness=-1\,  and 
Munich Center for \\ Quantum Science and Technology,\looseness=-2\, 
Ludwig-Maximilians-Universit\"at M\"unchen, 80333 Munich, Germany}
\author{Andreas Weichselbaum}
\affiliation{Condensed Matter Physics and Materials Science Department,\looseness=-1\,  
Brookhaven National Laboratory, Upton, NY 11973, USA}
\affiliation{Arnold Sommerfeld Center for Theoretical Physics, 
Center for NanoScience,\looseness=-1\,  and 
Munich Center for \\ Quantum Science and Technology,\looseness=-2\, 
Ludwig-Maximilians-Universit\"at M\"unchen, 80333 Munich, Germany}
\author{Gabriel Kotliar}
\affiliation{Condensed Matter Physics and Materials Science Department,\looseness=-1\,  
Brookhaven National Laboratory, Upton, NY 11973, USA}
\affiliation{Department of Physics and Astronomy, Rutgers University, Piscataway, NJ 08854, USA}
\author{Jan von Delft}
\affiliation{Arnold Sommerfeld Center for Theoretical Physics, 
Center for NanoScience,\looseness=-1\,  and 
Munich Center for \\ Quantum Science and Technology,\looseness=-2\, 
Ludwig-Maximilians-Universit\"at M\"unchen, 80333 Munich, Germany}

\date{September 27, 2019}

\begin{abstract}
Orbital differentiation is a common theme in multiorbital systems,
yet a complete understanding of it is still missing.
Here, we consider a minimal model for orbital differentiation in Hund metals with a highly accurate method:
We use the numerical renormalization group as a real-frequency impurity solver for a dynamical mean-field study of three-orbital Hubbard models, where a crystal field shifts one orbital in energy.
The individual phases are characterized with dynamic correlation functions and their relation to diverse Kondo temperatures.
Upon approaching the orbital-selective Mott transition,
we find a strongly suppressed spin coherence scale and uncover the emergence of a singular Fermi liquid and interband doublon-holon excitations.
Our theory describes the diverse polarization-driven phenomena in the $t_{2g}$ bands of materials such as ruthenates and iron-based superconductors, and our methodological advances pave the way toward real-frequency analyses of strongly correlated materials.
\end{abstract}

\maketitle

\section{Introduction}
The discovery of superconductivity in the iron pnictides and
chalcogenides \cite{Kamihara2006,Kamihara2008} (FeSCs) has led to renewed
interest in multiorbital systems. Both theoretical and experimental studies of these systems have
uncovered the remarkable phenomenon of orbital differentiation: In an
almost degenerate manifold of $d$ states, some orbitals are markedly
more correlated than others. For instance, in FeSe$_x$Te$_{1-x}$
\cite{Liu2015}, LiFeAs \cite{Miao2016}, and
K$_{0.76}$Fe$_{1.72}$Se$_2$ \cite{Yi2015}, among the $t_{2g}$
states, only the $xy$ orbital disappears from
photoemission
spectra as temperature is raised.  Orbital differentiation is also
seen in tunneling experiments \cite{Sprau2017} and is a key ingredient in theoretical
frameworks to describe FeSCs \cite{Yin2011,deMedici2014,deMedici2014a}. It is not
unique to the FeSCs; it has further been documented in the
ruthenates \cite{Sutter2019} and likely takes place in all
Hund metals \cite{Haule2009,Georges2013}.
An extreme form of orbital differentiation is the orbital-selective
Mott transition (OSMT) \cite{Anisimov2002}, where 
some orbitals become insulating, while others
remain metallic. 
Despite its importance, the OSMT in 
three-band systems 
has not yet been
systematically investigated with a controlled method enabling
access to low temperatures, where Fermi liquids form. 
Controversial questions include:
For a given sign of
crystal-field splitting, which orbitals localize? 
Is the OSMT of first or second order? 
Do correlations enhance or reduce orbital
polarization as one approaches the OSMT?
Is it true that quenching of orbital fluctuations makes the orbitals behave independently?
Do the 
itinerant electrons in the OSM phase
(OSMP) form a Fermi liquid?
Finally, how are the 
precursors of the OSMT related to the physics of Hund metals?
In this paper, 
we use a minimal model (see motivation below) for orbital differentiation 
in Hund metals
to answer these questions in a unified picture.
Our conceptual arguments are supported by a numerical method of 
unprecedented accuracy: We
use the numerical renormalization group (NRG) \cite{Bulla2008} as a real-frequency impurity solver
for dynamical mean-field theory (DMFT) \cite{Georges1996}, extending the tools of Ref.~\onlinecite{Stadler2015} from full SU(3) to reduced orbital symmetry.
Whereas different bandwidths directly lead to different effective interaction strengths among the orbitals (as extensively studied for two-orbital models; see, e.g., \cite{Inaba2007} for a list of references), 
we focus here on the more intricate case 
where a crystal field shifts one orbital in energy 
w.r.t.\ two degenerate orbitals
\cite{deMedici2009,Werner2009,Kita2011,Huang2012,Wang2016}.
Thereby, we can isolate polarization 
effects and drive the system through band+Mott insulating, metallic, and OSM phases, reminiscent of Ca$_2$RuO$_4$ \cite{Anisimov2002}, Sr$_2$RuO$_4$ \cite{Mravlje2011}, and FeSCs, respectively.
Theoretically, the OSMP has been under debate both w.r.t.\ the precise
form of the (conducting) self-energy
\cite{Biermann2005,Werner2006,deMedici2009,deMedici2011a,Huang2012}
and w.r.t.\ subpeaks in the insulating spectral function
\cite{deMedici2005,Ferrero2005,Kita2011,deMedici2011a}. 
Whereas previous studies were limited by finite-size effects of exact
diagonalization or finite temperature in Monte Carlo data
(requiring analytic continuation), our NRG results yield
conclusive numerical evidence down to the lowest energy scales.
We give a detailed phase
diagram 
including coexistence regimes (lacking hitherto)
and characterize the
individual 
phases with real-frequency properties and their relation
to Kondo temperatures spanning several orders of magnitude.
Upon approaching the OSMT, we find a 
strongly suppressed spin coherence scale
and uncover the emergence of a  singular Fermi liquid
\cite{Coleman2003,Mehta2005,Koller2005,Biermann2005,Greger2013} 
and interband doublon-holon excitations \cite{Yee2010,Haule2010,Nunez2018,Komijani2019}
(both of which were previously realized only separately and in two-orbital models).
\section{Model and method}
The Hamiltonian of our three-orbital Hubbard model is given by
\begin{equation*}
\hat{H} = 
-t \!\! \sum_{\langle i j \rangle m\sigma} \hat{d}^{\dag}_{im\sigma} \hat{d}_{jm\sigma}
+ \sum_i \hat{H}_{\textrm{int}} [\hat{d}_{im\sigma}]
+ \sum_{im} \epsilon_m \hat{n}_{im}
\EC
\end{equation*}
where $\hat{d}^{\dag}_{im\sigma}$ creates an electron on lattice site $i$ in orbital $m \in \{1,2,3\}$ with spin $\sigma \in \{ \uparrow, \downarrow \}$.
The first term describes nearest-neighbor hopping within each orbital on the lattice of uniform amplitude $t=1$, which thus sets the unit of energy.
As local interaction, we use the following ``minimal rotationally invariant'' form \cite{Dworin1970,Georges2013,Stadler2015,Horvat2016},
\begin{equation*}
\hat{H}_{\textrm{int}} [\hat{d}_{m\sigma}] = 
\tfrac{3}{4} J \hat{n}
+ \tfrac{1}{2} \big( U - \tfrac{3}{2} J \big ) \hat{n} \big( \hat{n} - 1 \big)
- J \hat{\vec{S}}^2
\ED
\end{equation*}
Here, 
$\hat{\vec{S}} =  \sum_m \hat{\vec{S}}_m$ 
is the total spin operator; 
$\hat{n} = \sum_m \hat{n}_m$,
$\hat{n}_m = \sum_{\sigma} \hat{n}_{m\sigma}$, and
$\hat{n}_{m\sigma} = \hat{d}^{\dag}_{m\sigma} \hat{d}_{m\sigma}$
are number operators
with expectation values 
$n$, $n_m$, and $n_{m\sigma}$, respectively.
This interaction yields an 
intraorbital Coulomb interaction of size $U$, 
interorbital Coulomb interactions of size 
$U-J$ and $U-2J$ for opposite and equal spins, respectively,
and a spin-flip term proportional to $J$
[cf.\ Eq.~\eqref{eq:Hint}]. 
With only two
parameters, it exhibits the full SU(3) symmetry, as opposed to the SO(3) symmetry of the usual Hubbard--Kanamori Hamiltonian \cite{Kanamori1963,Georges2013}.
We mostly fix these parameters to
$U=6$ and $J=1$.
Our only source of orbital differentiation 
comes from the last term in $\hat{H}$
via the crystal-field splitting $\Delta$,
defined as relative shift among the on-site energies
(cf.\ Fig.~\ref{fig:levels}):
$\epsilon_1 - \Delta = \epsilon_2 = \epsilon_3 \equiv \epsilon_{23}$.
(The index ``$23$'' indicates shared  properties of the degenerate
doublet, e.g., $n_{23} \equiv n_2 = n_3$.)
The overall shift of $\epsilon_m$ is determined by the average filling $n=2$,
taken one away from half filling as characteristic for Hund metals.
Note that, for $J$ to act nontrivially, this setting requires at least three orbitals.
While the effect of $\Delta$ in uncorrelated systems is rather straightforward, 
the interplay of $\Delta$ with $U$ and especially $J$ in Hund metals leads to
intriguing phenomena.
Within the DMFT approximation,
the lattice Hamiltonian is mapped to an impurity problem with self-consistently determined hybridization \cite{Georges1996}. We use a semicircular lattice density of states (half-bandwidth 2), for convenience, and restrict ourselves to paramagnetic solutions at zero temperature ($T=10^{-8}$, in practice).
The impurity problem is solved on the real-frequency axis 
by means of the full-density matrix \cite{Weichselbaum2007} NRG.
The numerical challenge of three orbitals with reduced symmetry is overcome by 
interleaving the Wilson chains \cite{Mitchell2014,Stadler2016}
of the 1-orbital and 23-doublet, while
fully exploiting the remaining $\textrm{SU}(2)_{\textrm{spin}} \otimes \textrm{U}(1)_{\textrm{charge,1}} \otimes \textrm{U}(1)_{\textrm{charge,23}} \otimes \textrm{SU}(2)_{\textrm{orbital,23}}$ symmetry,
using the QSpace tensor library \cite{Weichselbaum2012,Weichselbaum2012a}.
We set the overall discretization parameter to $\Lambda=6$ 
and keep up to 30000 multiplets ($\sim 2.5\cdot 10^5$ states) during the iterative diagonalization.
While NRG can famously resolve arbitrarily small energy scales very accurately, we also obtain a sufficiently accurate resolution at high energies via adaptive broadening \cite{Lee2016,Lee2017} of the discrete spectral data obtained for two different $z$ shifts \cite{Zitko2009}.
As dynamic correlation functions, we compute the impurity self-energy $\Sigma$ \cite{Bulla1998}, also used to extract the DMFT local spectral function $\mathcal{A}$, as well as spin and orbital susceptibilities $\chi=\chi'-i\pi\chi''$, 
defined in Appendix \ref{sec:susceptibilities}.
\section{Crystal-field splitting}
\begin{figure}[t]
\includegraphics[width=.48\textwidth]{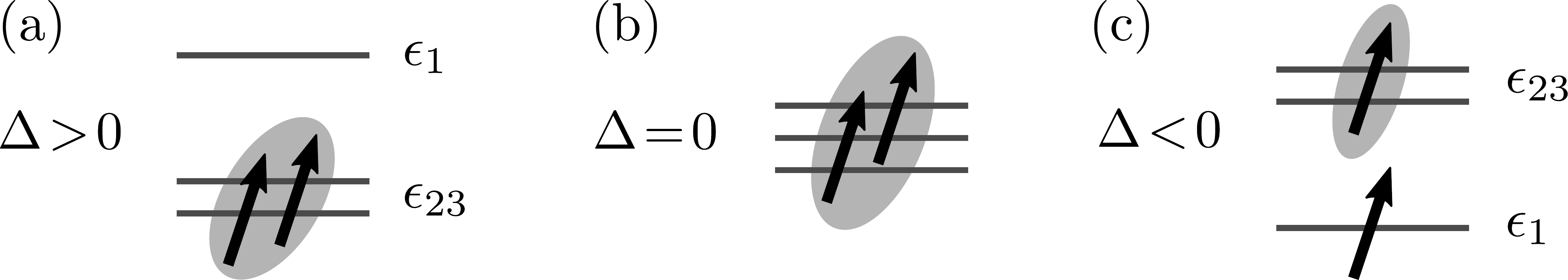}
\caption{%
Illustration of the on-site energies $\epsilon_1-\Delta=\epsilon_2=\epsilon_3$ and impurity occupations. Due to Hund's coupling, spins are aligned; shaded arrows symbolize a symmetric distribution among the degenerate orbitals.
The different phases portrayed are (a) a band+Mott insulator for large, positive $\Delta$,
(b) an orbitally symmetric metal for vanishing $\Delta$, 
and (c) the OSMP for large, negative $\Delta$,
yet $|\Delta|\lesssim 2J$.
After a particle-hole transformation and the identification $1\leftrightarrow xy$, $23\leftrightarrow xz/yz$, (a) and (b) mimic properties of the $t_{2g}$ orbitals of Ca$_2$RuO$_4$ and Sr$_2$RuO$_4$, respectively; with a half-filled $xy$ orbital and further metallic ones, (c) resembles the situation in FeSCs.%
}
\label{fig:levels}
\end{figure}

\begin{figure}[t]
\includegraphics[width=.49\textwidth]{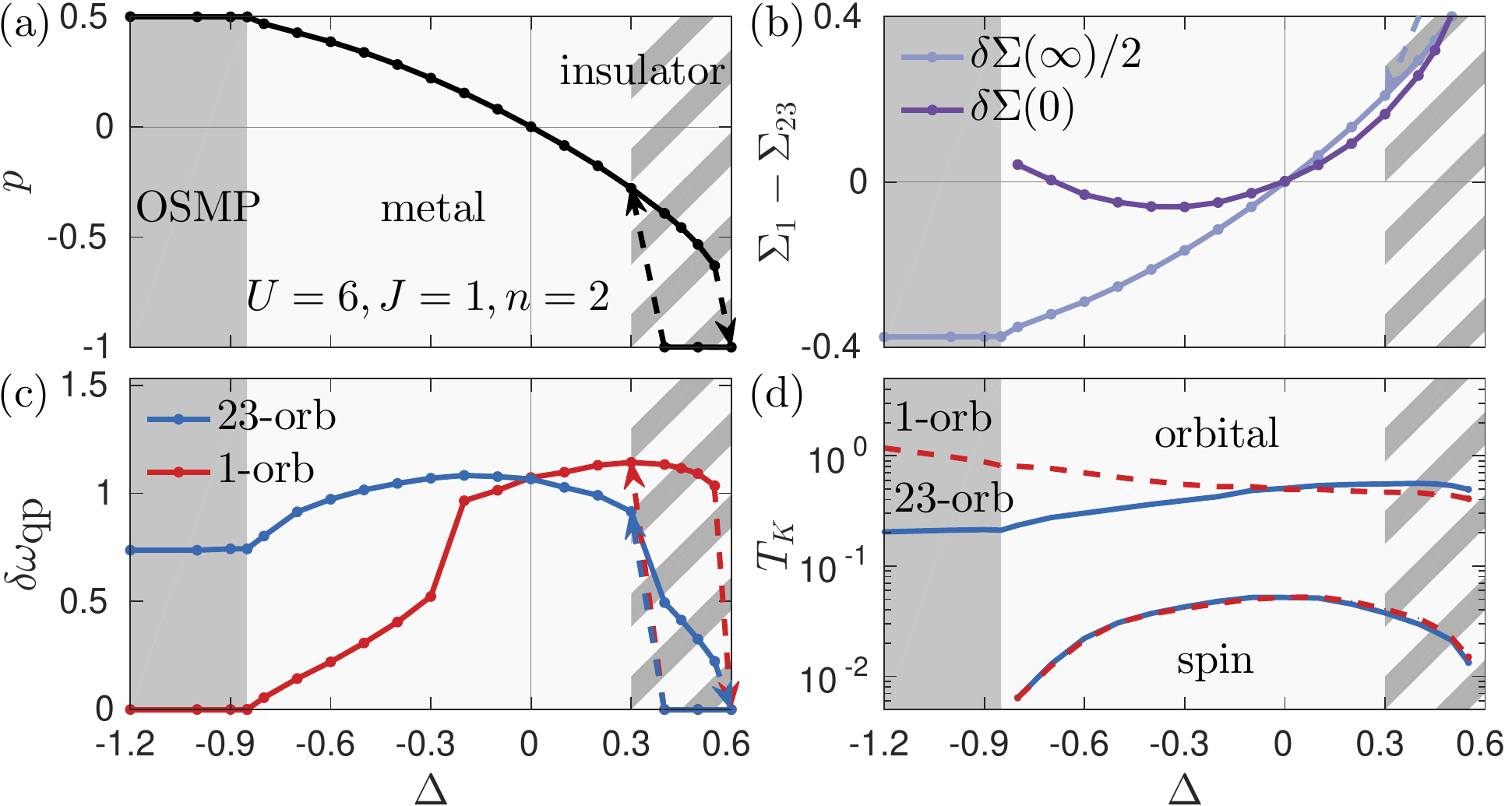}
\caption{%
Phase diagram for varying $\Delta$.
(a) The orbital polarization, $p=n_1 -n_{23}$, directly hints at the different phases 
portrayed in Fig.~\ref{fig:levels}.
We find coexisting solutions for $\Delta \in [0.3,0.6]$
but no hysteresis between metal and OSMP.
(b) The self-energy difference, $\delta\Sigma = \textrm{Re}\Sigma_1-\textrm{Re}\Sigma_{23}$, adds to a renormalized $\Delta$. Whereas $\delta\Sigma$ increases with increasing $\Delta>0$ at both $\omega \in \{\infty,0\}$, the $\delta\Sigma(0)$ curve (only shown for metallic solutions) bends upward for $\Delta<-0.3$, thereby counteracting the splitting.
(c) The full width at half maximum of the quasiparticle peak, $\delta\omega_{\textrm{qp}}$, confirms the metallic vs.\ insulating character. 
In the coexistence regime, either $\delta\omega_{\textrm{qp}}=0$ or $\delta\omega_{\textrm{qp}}>0$ for all orbitals alike.
(d) The orbital and spin Kondo temperatures are clearly separated 
($T_K\orb \simeq 0.5$, $T_K\sp \simeq 0.05$ at $\Delta=0$).
Strikingly, $T_K\sp$ strongly decreases with increasing $|\Delta|$ and vanishes altogether in the OSMP 
(out of range on the log scale).%
}
\label{fig:phase_diagr}
\end{figure}

As we tune $\Delta$,
the system undergoes (for suitable interaction strength) several phase transitions.
The nature of the different phases can be easily understood by looking at the occupations in the atomic limit (Fig.~\ref{fig:levels}) \cite{Werner2009,Huang2012}:
For large $\Delta>0$, the 1-orbital has highest energy; both electrons reside in the half-filled 23-doublet and are likely to form a Mott insulator \cite{deMedici2011}.
For the symmetric model at $\Delta=0$, the two electrons are equally distributed among the three degenerate orbitals with occupation $n_m=2/3$ each, giving rise to metallic behavior (for not too strong interaction).
Finally, for large $\Delta<0$, the filling of the lowest orbital is eventually increased up to half filling, $n_1 = 1$, and the remaining electron occupies the quarter-filled 23-doublet. For intermediate interaction strengths 
\footnote{For $\Delta<0$, the Mott transition of the half-filled 1-orbital depends mainly on $U$, with a critical $U_{c2} \approx 6$ similar to the single-orbital case \cite{Georges1996}. The Mott transition in the quarter-filled, 23-doublet requires much stronger interaction \cite{Gorelik2013,Lee2018}.
Hence, our choice $U=6$ and $J=1$ is close to the minimal interaction strength required for the OSMP.}, 
the half-filled 1-orbital is Mott-insulating while the quarter-filled 23-doublet remains metallic, thereby realizing an OSMP.
By decreasing $\Delta$ even further, one reenters a metallic ($1 < n_1 < 2$) and ultimately a band-insulating phase ($n_1=2$).
\nocite{Gorelik2013,Lee2018}
These considerations anticipate the mechanism driving the phase transitions
\cite{deMedici2009,Werner2009,Kita2011,Huang2012,Wang2016,Steinbauer2019}:
$\Delta$ primarily induces orbital polarization; i.e., it changes the relative \textit{filling} of the orbitals. 
Starting from the orbitally symmetric, metallic phase, the different orbitals can become band-insulating or undergo a \textit{filling-driven} Mott transition. 
If there are partially filled orbitals of different occupations and/or degeneracies, as 
in Fig.~\ref{fig:levels}(c), this leads to different critical interaction strengths for the Mott transition, and an OSMP can be realized.
We now investigate the precise nature of these phase transitions as function of $\Delta$ for fixed $U$, $J$, $n$.
Figure~\ref{fig:phase_diagr}(a) shows the orbital polarization, $p=n_1 -n_{23}$.
Starting from the symmetric case 
($\Delta=0$, $p=0$)
and increasing $\Delta$, $p$ decreases to its minimum $p=-1$
[cf.\ Fig.~\ref{fig:levels}(a)].
For large $\Delta>0$, we observe a coexistence region
when approaching $\Delta$ from below or above,
giving rise to the definitions $\Delta^{\textrm{pos}}_{c1} \simeq 0.3$, $\Delta^{\textrm{pos}}_{c2} \simeq 0.6$.
If we decrease $\Delta$ starting from $\Delta=0$, $p$ increases until it saturates for $\Delta \leq \Delta^{\textrm{neg}}_c \simeq -0.85$ at $p=0.5$. 
This regime 
constitutes the OSMP, for which we find \textit{no} hysteresis w.r.t.\ $\Delta$.
Clearly, the $\Delta$-driven OSMT is much more second-order-like than the ordinary Mott transition at $\Delta>0$.
We also note that, while $p$ appears differentiable at the OSMT, $\textrm{Var}(\hat{p})$ exhibits a kink
[cf.\ Fig.~\ref{fig:app_stat}(a)].
The OSMP is stable from $\Delta^{\textrm{neg}}_c$ down to $\Delta \simeq -1.5$, where one enters a strongly polarized ($p>0.5$) metallic phase (not shown).
To address the effect of correlations on orbital differentiation, we examine the difference in the real part of the self-energies, $\delta \Sigma = \textrm{Re}\Sigma_1 - \textrm{Re}\Sigma_{23}$, which adds to a renormalized crystal field \cite{Kita2011}, $\Delta+\delta\Sigma$ [cf.\ also Fig.~\ref{fig:app_stat}(b)].
The overall shift of the self-energies is given by the Hartree part, $\Sigma_{\textrm{H}}=\Sigma(\omega=\infty)$, which can directly be calculated:
\begin{equation*}
\Sigma_{\textrm{H},m\sigma} = 
U n_{m\bar{\sigma}} +
\sum_{m' \neq m} \Big[
(U-J)  n_{m' \bar{\sigma}} +
(U-2J) n_{m' \sigma}
\Big]
\ED
\end{equation*}
The difference, $\delta\Sigma_{\textrm{H}}=-(U-3J)p/2$,
increases monotonically with $\Delta$ (via $p$) for $U-3J>0$, such that
interactions overall enhance orbital differentiation \cite{Georges2013}.
However, the renormalization of $\Delta$ at low energies
must be determined numerically.
Figure~\ref{fig:phase_diagr}(b) displays $\delta\Sigma$ at $\omega \in \{0,\infty\}$:
$\delta\Sigma(0)$ is smaller in magnitude than $\delta\Sigma_{\textrm{H}}$ (plot shows $\delta\Sigma_{\textrm{H}}/2$) and increases monotonically with $\Delta$ only for $\Delta>-0.3$. For $\Delta<-0.3$, $\delta\Sigma(0)$ bends upward and eventually increases with decreasing $\Delta$, thereby counteracting the splitting.
Next, Fig.~\ref{fig:phase_diagr}(c) shows the 
width of the quasiparticle peak, $\delta\omega_{\textrm{qp}}$, of the spectral function (cf.\ Fig.~\ref{fig:met_spf}) to confirm the conducting vs.\ insulating character of the different phases.
For positive and negative $\Delta$, we indeed find that the 23- and 1-orbital(s), respectively, undergo a Mott transition, with gradually decreasing $\delta\omega_{\textrm{qp}}$.
The sharp decline in $\delta\omega_{\textrm{qp}}$ around $|\Delta| \sim 0.3$ corresponds to the formation of a subpeak (see below).
For $\Delta>0$, the 1-orbital shows a slight increase of $\delta\omega_{\textrm{qp}}$ and eventually becomes band-insulating, while,
for $\Delta<0$, $\delta\omega_{\textrm{qp}}$ of the 23-orbitals decreases until it saturates in the OSMP.
Note that the quasiparticle weight, $Z_m=[1-\partial_{\omega}\textrm{Re}\Sigma_m(0)]^{-1}$,
often used to describe the single-orbital Mott transition, is not ideal to characterize the full range of orbital differentiation:
For $\Delta>0$, when the 1-orbital gets emptied out, $Z_1$ increases although the whole quasiparticle peak gradually disappears;
for large $\Delta<0$, $Z_1$ of the insulating 1-orbital does not vanish throughout the OSMP,
yet $Z_{23}=0$ in the metallic 23-orbitals, as further explained below.
We complete our phase diagram by showing in Fig.~\ref{fig:phase_diagr}(d)
the $\Delta$-dependence of Kondo temperatures, defined as the energy scale at which the corresponding susceptibility, $\chi''$, is maximal [cf.\ Fig.~\ref{fig:met_spf}(d)].
As typical for Hund metals \cite{Georges2013,Stadler2015}, we observe spin--orbital separation in terms of Kondo scales:
orbital fluctuations 
are screened at much higher energies than spin fluctuations ($T_K^{\textrm{orb}} \gg T_K^{\textrm{sp}}$).
While $T_{K,23}\orb$ characterizes orbital fluctuations within the 23-doublet,
$T_{K,1}\orb$ describes those between the (separated) 1-orbital and the 23-doublet
[cf.\ Eq.~\eqref{eq:chiorb}]
and reduces to the bare energy scale $\sim|\Delta|$ for large splitting.
At sizable $J$, both orbitals have the same $T_K\sp$ \cite{Greger2013a},
and, strikingly, $T_K\sp$ strongly decreases with increasing $|\Delta|$.
\begin{figure}[t!]
\includegraphics[width=.48\textwidth]{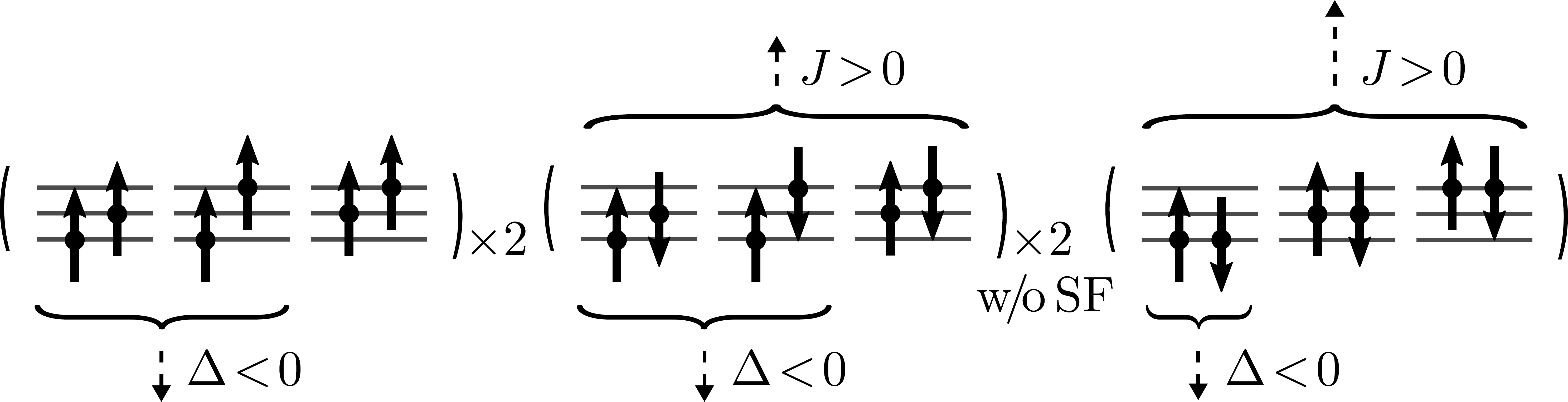}
\caption{%
Illustration of all 15 different impurity states for $n=2$ in the $\hat{n}_m$, $\hat{S}_m^z$ basis.
Finite $J$ and $\Delta$ yield a relative shift in the eigenenergies (dashed arrows) 
and thus split the $J=0=\Delta$ ground-state manifold.
The states in the middle are eigenstates of the impurity Hamiltonian 
only without spin-flip (SF) terms,
where Hund's coupling merely shifts the density-density interactions by
$J$ and $2J$ [cf.\ Eq.~\eqref{eq:Hint}];
with SU(2) spin symmetry, they form singlet and triplet combinations.
Subscripts $\times 2$ indicate that the number of states is counted twice due to spin 
degeneracy.
Without SF terms, 
the ground-state degeneracy of 15 at $J=0=\Delta$ is reduced to
6 at $J>0$, $\Delta=0$ and to 4 at $J>0$, $\Delta<0$.
Including SF terms, these are 15, 9, and 6.%
}
\label{fig:gs_deg}
\end{figure}
This can be understood as follows:
It is well-known that finite $J$ decreases $T_K^{\textrm{sp}}$ \cite{Okada1973,Georges2013,Stadler2018}, as it splits the impurity ground-state manifold.
Intuitively, a smaller ground-state degeneracy implies a reduced effective hybridization and thus a reduced Kondo temperature.
For $J>0$ and finite $\Delta$, the ground-state degeneracy is reduced even further, particularly for $\Delta<0$; 
see Fig.~\ref{fig:gs_deg}. 
Moreover, the DMFT self-consistency suppresses the low-energy hybridization strength of the orbital approaching the Mott transition. In the OSMP, $\mathcal{A}_1(0)$ and $T_K\sp$ eventually
vanish altogether.
\section{Metallic spectrum}
\begin{figure}[t]
\includegraphics[width=.48\textwidth]{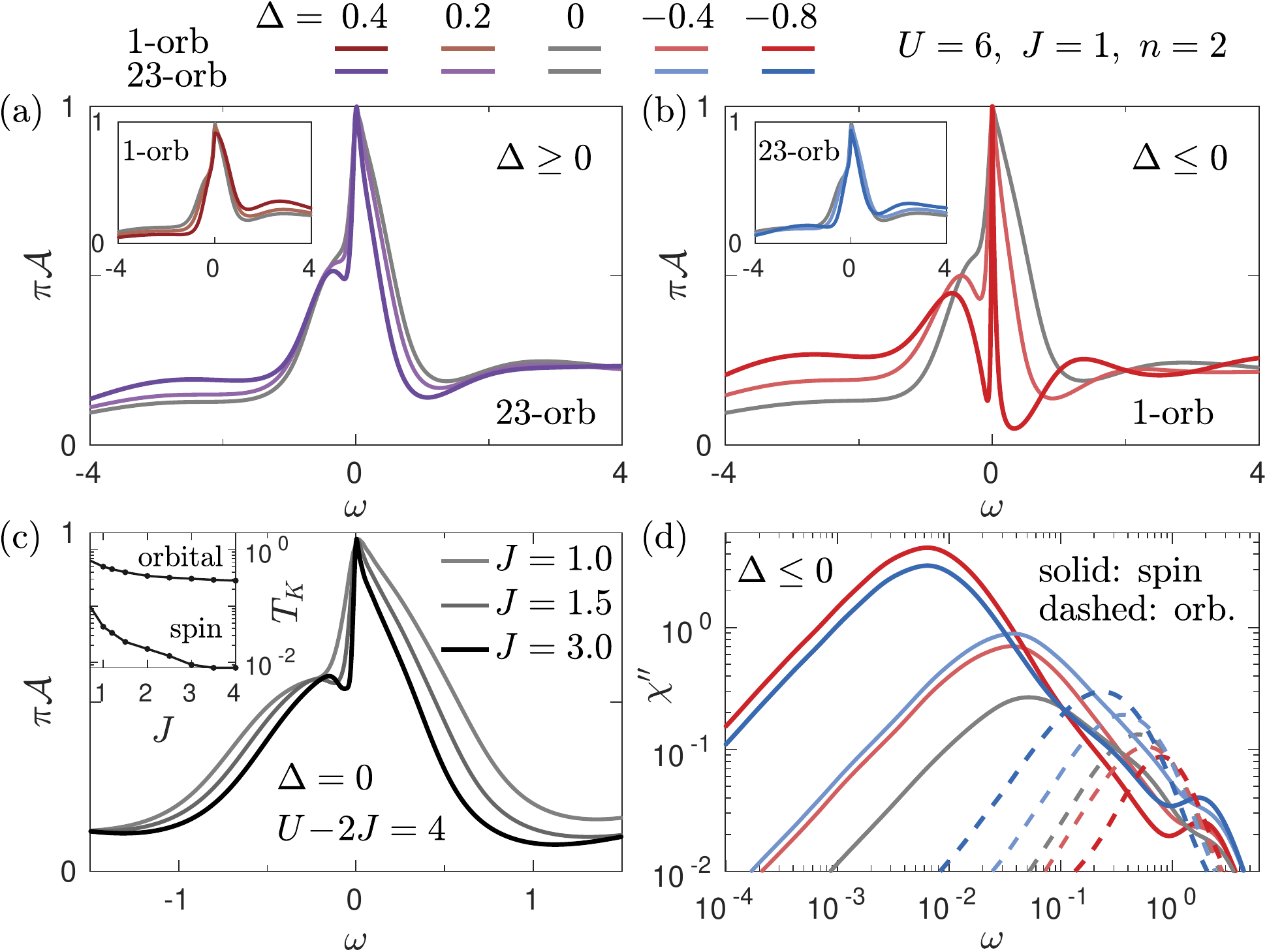}
\caption{%
(a,b) Spectral functions in the metallic phase for
the orbitals approaching a Mott transition (main panels) and the remaining ones (insets).
(b) Decreasing $\Delta$ sharpens the quasiparticle peak (reduced $T_K\sp$), 
destroys the orbital resonance 
if $|\Delta| \gtrsim T_K\orb(\Delta=0)/2$, 
and generates interband doublon-holon subpeaks.
(c) Spectral functions and Kondo temperatures (inset) in the orbitally symmetric case 
for increasing $J$ and fixed $U-2J$. 
(d) Spin (solid lines) and orbital (dashed) susceptibilities corresponding to (b). 
(For $\Delta\neq 0$, we 
plot $\chi\sp_1$ and $4\chi\sp_{23}$ to have the two curves for each $\Delta$ closer together.)%
}
\label{fig:met_spf}
\end{figure}

\begin{figure}[t!]
\includegraphics[width=.48\textwidth]{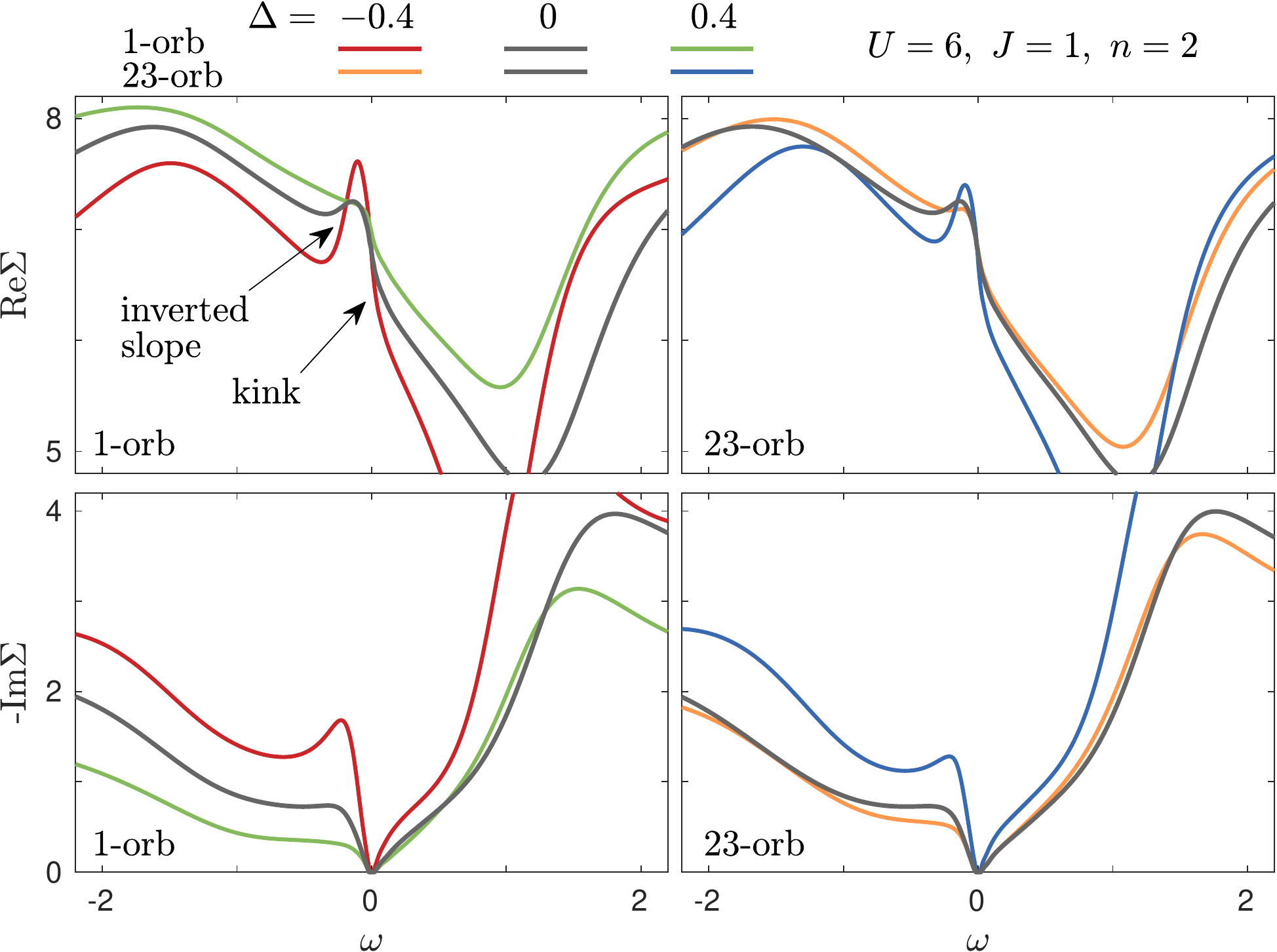}
\caption{%
Metallic self-energies for all orbitals, for different $\Delta$. The characteristic features,
such as an inverted slope and a kink,
already present at $\Delta=0$,
are enhanced as the orbital becomes more correlated, 
induced by proximity to half filling:
$n_1$ ($n_{23}$) approaches $1$ with increasing (decreasing) $\Delta$.%
}
\label{fig:SE_met}
\end{figure}

Let us now examine in detail how the spectral functions change with $\Delta$
in the metallic phase. Figures \ref{fig:met_spf}(a,b) show that, for both positive and negative $\Delta$, the most important change with stronger correlations occurs in the orbital(s) 
approaching a Mott transition (main panels). 
The other orbitals (insets) 
mostly adjust the spectral weight. 
At $\Delta=0$ [gray lines in Figs.~\ref{fig:met_spf}(a--c)], the spectral functions exhibit the typical shoulder in the quasiparticle (qp) peak \cite{Stadler2015,Stadler2018} (below half filling at $\omega<0$). In Ref.~\onlinecite{Stadler2018}, this has been explained as the combination of a sharp SU(2) spin Kondo resonance (``needle'' with width $\propto T_K^{\textrm{sp}}$) and a wider SU(3) orbital Kondo resonance (``base'' with width $\propto T_K^{\textrm{orb}}$).
If we first stay with the orbitally symmetric case [Fig.~\ref{fig:met_spf}(c)] and use $J$ and $E_{\textrm{at}} = U-2J$ as 
tuning parameters \cite{Stadler2018}, we can reduce $T_K^{\textrm{sp}}$ by increasing $J$ while only mildly affecting $T_K^{\textrm{orb}}$. As a consequence, the needle sharpens while the wide base remains, revealing a subpeak.
Similarly, increasing $|\Delta|$ 
drastically decreases $T_K^{\textrm{sp}}$ [Figs.~\ref{fig:phase_diagr}(d), \ref{fig:met_spf}(d)] 
and causes a 
thin qp needle.
Additionally, finite $\Delta$, which acts in orbital space similarly to a magnetic field in spin space,
splits the qp base. For 
$|\Delta| \gtrsim T_K\orb$, the orbital Kondo resonance is destroyed and subpeaks on both sides of $\omega=0$ remain.
In fact, the orbital-resonance shoulder is remarkably accurately centered at $-T_K\orb(\Delta\!=\!0)/2$ 
[Fig.~\ref{fig:met_spf}(c)], 
and crosses over to an interband doublon-holon excitation at $\Delta<0$ (see below)
for $|\Delta| \gtrsim T_K\orb(\Delta\!=\!0)/2$.
Note that the authors of Ref.~\onlinecite{Horvat2016} similarly marked strong influence of $J$ by $J \gtrsim T_K\orb(J\!=\!0)$.
Generally, finite $\Delta$ amplifies Hund-metal features in some orbitals
while suppressing them in others.
This is apparent in
spectral functions (Fig.~\ref{fig:met_spf}) as well as self-energies; 
see Fig.~\ref{fig:SE_met}.
For $\Delta=0$, we find the typical 
\cite{Iwasawa2005,Mravlje2011}
inverted slope in $\textrm{Re}\Sigma$ for small $\omega<0$
and kink in $\textrm{Re}\Sigma$ for small $\omega>0$ 
(with $\textrm{Im}\Sigma$ related by Kramers--Kronig transform). 
These features are enhanced as the orbital becomes more correlated, and suppressed as it becomes less correlated.
The degree of correlation follows from proximity to half filling: 
$n_1$ approaches $1$ as $\Delta$ decreases; $n_{23}$ approaches $1$ as $\Delta$ increases.
\section{OSMP}
\begin{figure}[t]
\includegraphics[width=.48\textwidth]{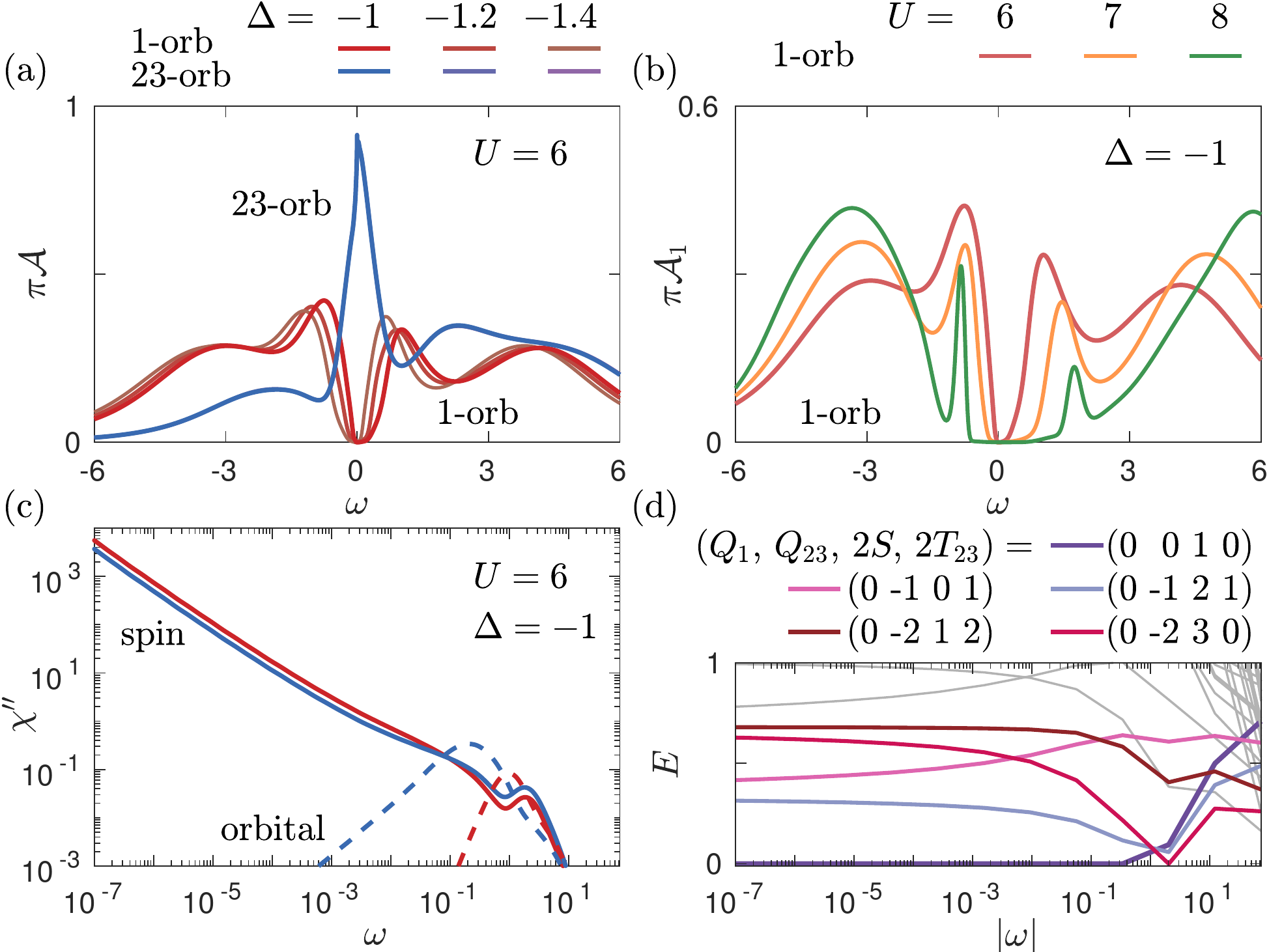}
\caption{%
Characterization of the OSMP. 
(a) Spectral functions showing the insulating and metallic character of the 1- and 23-orbital(s), respectively. Interband doublon-holon excitations are seen as subpeaks in $\mathcal{A}_1$, whose position shifts with $\Delta$, leading to a tilt of $\mathcal{A}_1$ around $\omega=0$;
the $\mathcal{A}_{23}$ curves all lie on top of each other.
(b) Close up of the insulating spectral function at variable $U$ (only in this panel), with $J/U=1/6$ fixed. 
Increasing $J$ shifts the right subpeak toward larger energies, and increasing $E_{\textrm{at}}=U-2J$ decreases the weight of the subpeaks by suppressing charge fluctuations in the $23$-doublet.
Both effects help to reveal a hard spectral gap.
(c) Diverging spin (solid lines) and regular orbital (dashed) susceptibilities. 
(We again plot $4\chi\sp_{23}$.)
(d) NRG flow diagram of the rescaled, lowest-lying energy levels at characteristic level spacing $\sim|\omega|$.
The legend provides charge $Q_m$, total spin $S$, and SU(2) orbital $T_{23}$ quantum numbers.
The ground state carries a residual spin $1/2$ since the contribution to the impurity spin from the insulating 1-orbital cannot be screened.
The SFL nature entails that the flow approaches the Fermi-liquid fixed point (where the first and second as well as third and fourth excitations become degenerate) only asymptotically.%
}
\label{fig:sfl_prop}
\end{figure}

For $\Delta \leq -0.85$, $T_K^{\textrm{sp}}$ and the qp needle vanish altogether; the 1-orbital becomes a Mott insulator while the 23-doublet retains spectral weight at $\omega=0$ [Fig.~\ref{fig:sfl_prop}(a)].
In the metallic orbitals, Luttinger pinning \cite{Mueller-Hartmann1989}
via the semicircular lattice density of states 
$\rho$, with $\mathcal{A}_{23}(0) = \rho(x_n)$ and 
$\int_{-\infty}^{x_n} \rho(x)\mathrm{d}x = n_{23,\sigma}$,
is fulfilled throughout
[leading to $\pi\mathcal{A}_{23}(0) \approx 0.91$ at quarter filling 
$n_{23,\sigma} = 1/4$].
Yet, the spectral function of the half-filled 1-orbital strongly differs from a single-orbital Mott insulator.
Next to the standard Hubbard bands,
charge fluctuations in the 23-doublet enable interband doublon-holon excitations 
(previously identified in a two-band DMFT+DMRG study \cite{Nunez2018};
cf.\ \cite{Yee2010,Haule2010} for experimental signatures)
in the insulating spectral function. Here, they occur at energies $\Delta$ and $\Delta+2J$, 
as derived in Appendix~\ref{sec:doublon-holon}.
These gap-filling states give $\mathcal{A}_1$ its soft form.
They are shifted with $\Delta$, leading to a ``tilt'' of $\mathcal{A}_1$ around $\omega=0$.
A hard gap is revealed when pushing the subpeaks apart (via $J$)
and decreasing their weight (via $E_{\textrm{at}}=U-2J$) by suppressing $23$-charge fluctuations 
[Fig.~\ref{fig:sfl_prop}(b)].
The subpeaks' distinct nature \cite{Lee2017,Lee2017a} is further underlined in plots of the momentum-resolved spectral function, shown in Appendix~\ref{sec:mom-res-A},
where one can also see how the widths of the 23-qp peak and 1-orbital subpeaks narrow together
with increasing $E_{\textrm{at}}$.
As the insulating 1-orbital does not contribute to spin screening, the OSMP inherits properties of an underscreened (spin) Kondo effect \cite{Greger2013},
as manifested in a divergent spin susceptibility [Fig.~\ref{fig:sfl_prop}(c)].
Within our DMFT description of the OSMP, the impurity electron in the 1-orbital 
and that in the 23-doublet form a combined spin 1, due to Hund's coupling. However, the 1-orbital 
hybridization ($\propto \mathcal{A}_1$)
has zero weight at low enough energies. Hence, given the diagonal hybridization, 
only the 23-contribution to the impurity spin can be screened, while its 1-orbital contribution remains unscreened.
The underscreened Kondo effect in turn leads to the singular Fermi-liquid (SFL) state
of the OSMP, as strikingly evident in the NRG flow diagram \cite{Bulla2008,Stadler2015,Stadler2018}:
Fig.~\ref{fig:sfl_prop}(d) shows that the rescaled, lowest-lying energy levels of the iteratively diagonalized Wilson chain reach the Fermi-liquid (FL) fixed point only asymptotically \cite{Mehta2005}.
\begin{figure}[t]
\includegraphics[width=.48\textwidth]{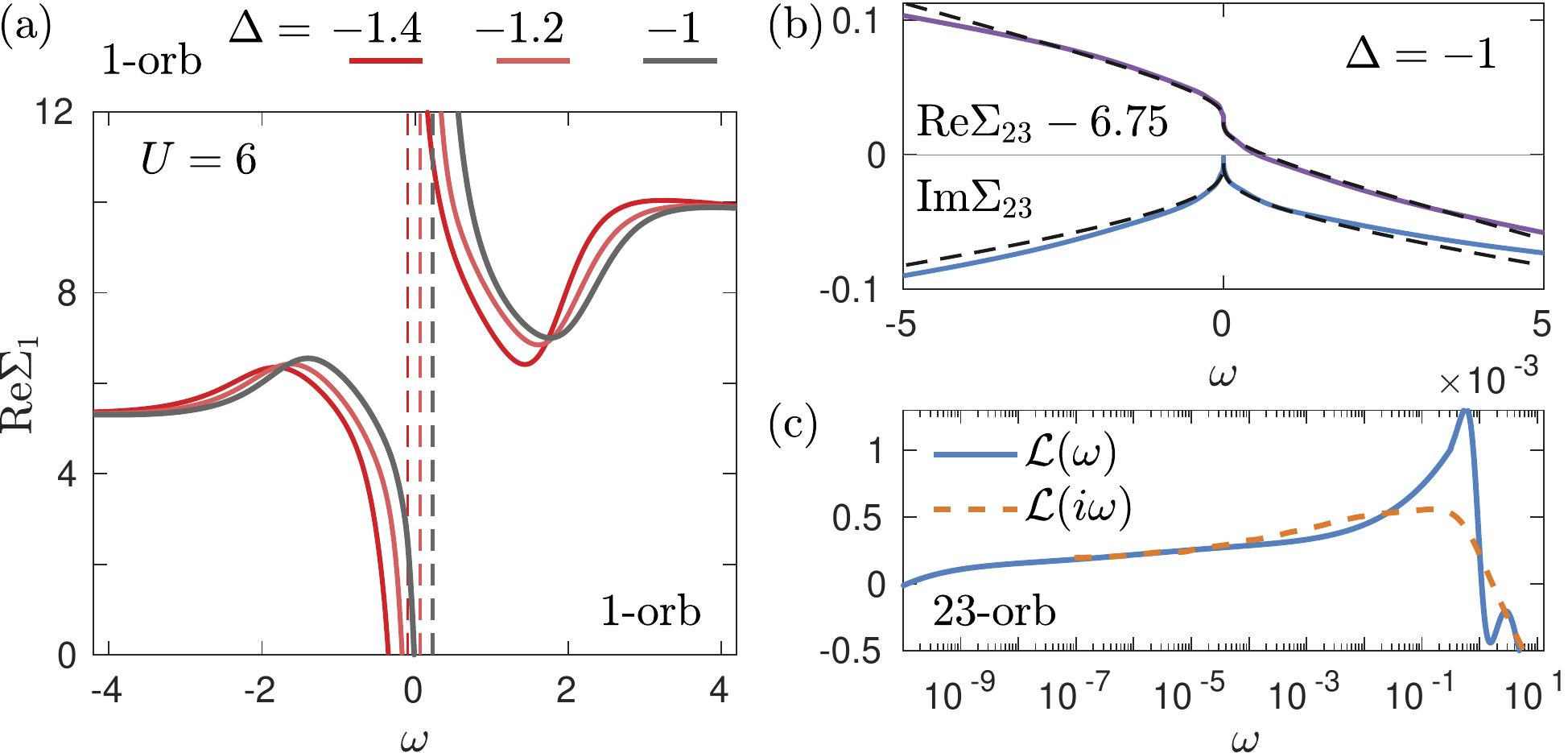}
\caption{%
Self-energies in the OSMP.
(a) Real part of the insulating 1-orbital self-energy.
Upon decreasing $\Delta$ in the OSMP, the position of the singularity 
in $\Sigma_1$ (marked by dashed lines) 
shifts through $\omega=0$.
(b) Low-energy zoom of the self-energy in the metallic 23-orbitals (solid lines) with fits (dashed) 
to the SFL logarithmic singularities.
(c) The logarithmic derivative $\mathcal{L}(z)$ of $-\mathrm{Im}\Sigma_{23}$ vanishes as $z\to 0$, 
providing additional confirmation of the logarithmic nature of the singularity.%
}
\label{fig:SE_OSMP}
\end{figure}
The self-energy of the insulating $1$-orbital diverges.
In Fig.~\ref{fig:SE_OSMP}(a), we see that the singularity of $\Sigma_1$ is not bound to $\omega=0$; instead, its position shifts with $\Delta$. 
This implies that $Z_1 = 1/(1 - \partial_\omega \textrm{Re}\Sigma_1(0))$ does not vanish throughout the OSMP and is thus not suited to mark the insulating character of the $1$-orbital in the OSMP.
A low-energy zoom of the self-energy 
in the metallic 23-orbitals [Fig.~\ref{fig:SE_OSMP}(b)] reveals strong deviations from the standard zero-temperature FL form, $\textrm{Re}\Sigma_{\textrm{FL}} = a + b \omega$ and $\textrm{Im}\Sigma_{\textrm{FL}} = -|c| \omega^2$.
Instead, it exhibits logarithmic singularities that can be well fitted [dashed lines in Fig.~\ref{fig:SE_OSMP}(b)] to the
SFL relations \cite{Biermann2005,Greger2013,Wright2011}
\begin{align*}
\textrm{Re} \Sigma_{\textrm{SFL}} & = \tilde{a} + \tilde{b} \textrm{ sgn}(\omega) \ln^{-3} | \omega / T^* | 
\EC \\
\textrm{Im} \Sigma_{\textrm{SFL}} & = -|\tilde{c}| \ln^{-2} | \omega / T^* |
\ED
\end{align*}
The logarithmic singularity in $\Sigma_{23}$ implies that $Z_{23}=0$ despite the conducting
character of the 23-orbitals with finite spectral weight at the Fermi level [Fig.~\ref{fig:sfl_prop}(a)].
To further scrutinize the singularity, we consider the logarithmic 
derivative of the imaginary part of $\Sigma_{23}$,
\begin{equation*}
\mathcal{L}(z) = \frac{\mathrm{d} \ln [ -\textrm{Im} \Sigma_{23}(z) ] }{\mathrm{d}\ln z}
\EC
\end{equation*}
both for real frequencies, $z=\omega+i0^+$ with $\omega \in \mathbb{R}$, and for imaginary frequencies, $z=i\omega \in (2\mathbb{Z}+1) i\pi T$.
This quantity is well suited to discriminate between singularities of logarithmic or fractional power-law type:
\begin{alignat*}{3}
-\textrm{Im}\Sigma(z) & = |c'| z^{\alpha}
&& \;\;\Rightarrow\;\;
\mathcal{L}(z) && = \alpha
, \\
-\textrm{Im}\Sigma(z) & = |\tilde{c}| \ln^{-2}(z/T^*) 
&& \;\;\Rightarrow\;\;
\mathcal{L}(z) && = -2 \ln^{-1}(z/T^*) 
\nonumber \\
& && && \xrightarrow[z \to 0]{} 0
.
\end{alignat*}%
In Fig.~\ref{fig:SE_OSMP}(c), we clearly see that $\mathcal{L}(0)=0$, confirming the logarithmic nature of the singularity.
Note that a smoothening postprocessing was used to suppress minor oscillations in very small values of $\textrm{Im}\Sigma$.
The imaginary-frequency data $\mathcal{L}(i\omega)$, available for $|i\omega| \geq \pi T$, perfectly match the low-frequency behavior but does not suffice to follow the decay up to $\mathcal{L}(0)=0$.
In fact, if the imaginary-frequency data were available only in a limited temperature range, as is the case in Monte Carlo studies,
say, $T \gtrsim 10^{-3}$ and $|i\omega| \gtrsim \pi \cdot 10^{-3}$, one might easily be tempted to conclude that $\mathcal{L}(i\omega)$ saturates at $\alpha \approx 0.5$.
\section{Conclusion}
We have 
shown that DMFT+NRG can be used to study 
three-orbital Hubbard models with reduced orbital symmetry, 
used this method to accurately describe polarization-driven phase transitions 
induced by a crystal field $\Delta$,
and uncovered the rich real-frequency structure inherent in the interplay of Hund-metal physics and orbital differentiation.
Our analysis leads to a conclusion of major conceptual significance:
The popular notion that orbital screening, facilitated by $J$, makes the orbitals behave almost independently 
\cite{deMedici2009,deMedici2011,deMedici2011a,Georges2013,deMedici2014,deMedici2014a,Fanfarillo2015,Sutter2019}
[as seen, e.g., in static correlations \cite{deMedici2009,deMedici2011a,Fanfarillo2015};
cf.\ also Fig.~\ref{fig:app_stat}(a)]
misses the importance of spin fluctuations.
It must be revised when looking at dynamic correlation functions, as
(i) a suppressed hybridization in one orbital suppresses the spin Kondo temperature of \textit{all} orbitals (at sizable $J$), 
(ii) charge fluctuations in some orbitals enable interband doublon-holon excitations \cite{Nunez2018} in the spectrum of other orbitals, 
and (iii) the presence of localized spins implies singular Fermi-liquid behavior of the remaining itinerant electrons \cite{Greger2013}.
With our methodological advances, NRG is ready to be used as a real-frequency impurity solver in a DFT+DMFT description of three-orbital materials with reduced orbital symmetry \cite{Kugler2019}.
Future studies should further investigate the stability of the OSMP against interorbital hopping \cite{Yu2017}.
\section*{Acknowledgments}
We thank A.~Georges and K.~M.~Stadler for fruitful discussions, and K.~Hallberg for a helpful correspondence.
FBK, S-SBL, and JvD were supported by the Deutsche Forschungsgemeinschaft under Germany's Excellence Strategy--EXC-2111--390814868; S-SBL further by Grant.\ No.\ LE3883/2-1.
FBK acknowledges funding from the research school IMPRS-QST. 
AW was funded by DOE-DE-SC0012704. GK was supported by NSF-DMR-1733071.
\appendix

\section{Additions to the phase diagram}
\begin{figure}[t!]
\includegraphics[width=.48\textwidth]{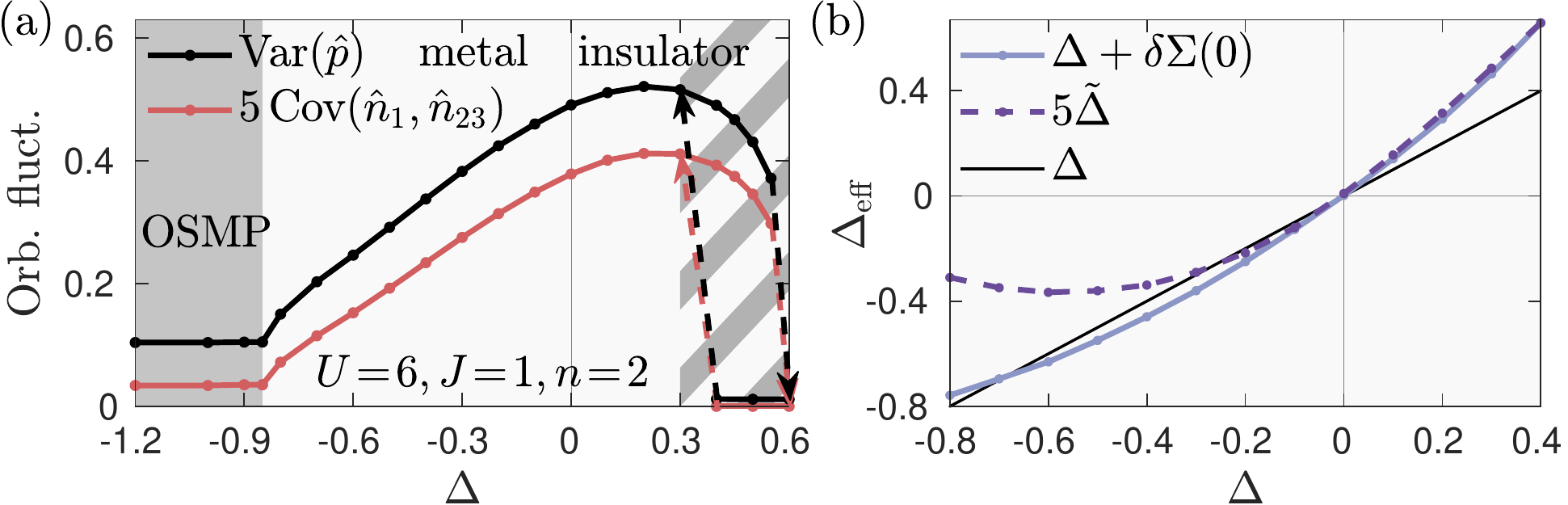}
\caption{%
Additions to the $\Delta$ phase diagram.
(a) 
$\textrm{Var}(\hat{p})=\langle \hat{p}^2 \rangle - \langle \hat{p} \rangle^2$ 
and
$\textrm{Cov}(\hat{n}_1,\hat{n}_{23})=|\langle \hat{n}_1 \hat{n}_{23} \rangle - \langle \hat{n}_1 \rangle \langle \hat{n}_{23} \rangle|$
exhibit a kink at the OSMT.
The latter shows that static, interband correlations are rather weak (plot shows $5\textrm{ Cov}$).
(b) Two different versions of an effective crystal field (shown only for metallic solutions), $\Delta+\delta\Sigma(0)$ as relevant for electronic degrees of freedom and $\tilde{\Delta} = Z_1 \cdot (\epsilon_1 + \Sigma_1(0)) - Z_2 \cdot (\epsilon_{23} + \Sigma_{23}(0))$ for quasiparticle excitations. Both show similar behavior: They depend monotonically on $\Delta$ in a region around $\Delta=0$ but bend upward for large, negative $\Delta$, counteracting the splitting.
}
\label{fig:app_stat}
\end{figure}

In the discussion of the phase diagram in Fig.~\ref{fig:phase_diagr}, 
we mentioned that the polarization 
$p = \langle \hat{p} \rangle$, with $\hat{p} = \hat{n}_1 - \hat{n}_{23}$ and $\hat{n}_{23} = (\hat{n}_2+\hat{n}_3)/2$,
varies with $\Delta$ in a differentiable way throughout the OSMT.
Regarding the nature of the phase transition, it is then interesting to note that 
$\textrm{Var}(\hat{p})=\langle \hat{p}^2 \rangle - \langle \hat{p} \rangle^2$ 
exhibits a kink at the OSMT [Fig.~\ref{fig:app_stat}(a)].
Further, we have elaborated on the intricate interorbital effects on dynamic correlation functions,
such as a strongly suppressed spin coherence scale, singular Fermi-liquid behavior, and interband doublon-holon excitations.
These effects are completely hidden when looking at static properties like the
interorbital correlator
$\textrm{Cov}(\hat{n}_1,\hat{n}_{23})=|\langle \hat{n}_1 \hat{n}_{23} \rangle - \langle \hat{n}_1 \rangle \langle \hat{n}_{23} \rangle|$, 
which, generally, is rather weak [Fig.~\ref{fig:app_stat}(b)] and has a kink at $\Delta_c^{\textrm{neg}}$ 
analogous to $\textrm{Var}(\hat{p})$
\cite{deMedici2009,deMedici2011a}.
To gauge the influence of correlations on orbital differentiation, we investigated
$\delta\Sigma(0) = \Sigma_1(0)-\Sigma_{23}(0)$,
which contributes to a renormalized crystal field, $\Delta+\delta\Sigma(0)$, for electronic degrees of freedom.
An alternative definition for an effective crystal field, $\Delta_{\textrm{eff}}$, is given by $\tilde{\Delta} = Z_1 \cdot  (\epsilon_1 + \Sigma_1(0)) - Z_2 \cdot (\epsilon_{23} + \Sigma_{23}(0))$, which constitutes a splitting for quasiparticle excitations 
\cite{Kita2011}.
Figure~\ref{fig:app_stat}(b) shows that both variants of $\Delta_{\textrm{eff}}$ vary similarly with $\Delta$:
In a region around $\Delta=0$, the self-energy difference $\delta\Sigma(0)$ increases the magnitude of $\Delta_{\textrm{eff}}$, i.e., $\delta\Sigma(0)>0$ for $\Delta>0$ and $\delta\Sigma(0)<0$ for moderate $\Delta<0$. However, for large, negative $\Delta$, we find that $\delta\Sigma(0)>0$ for $\Delta<0$, thus decreasing $|\Delta_{\textrm{eff}}|$.
The quasiparticle effective crystal field, $\tilde{\Delta}$, is much smaller in magnitude than the bare crystal field, but, nonetheless, shows a trend similar to that of $\Delta+\delta\Sigma(0)$: It depends monotonically on $\Delta$ in a region around $\Delta=0$ but bends upward for large, negative $\Delta$, thereby counteracting the splitting.
\begin{figure*}[t!]
\includegraphics[width=.99\textwidth]{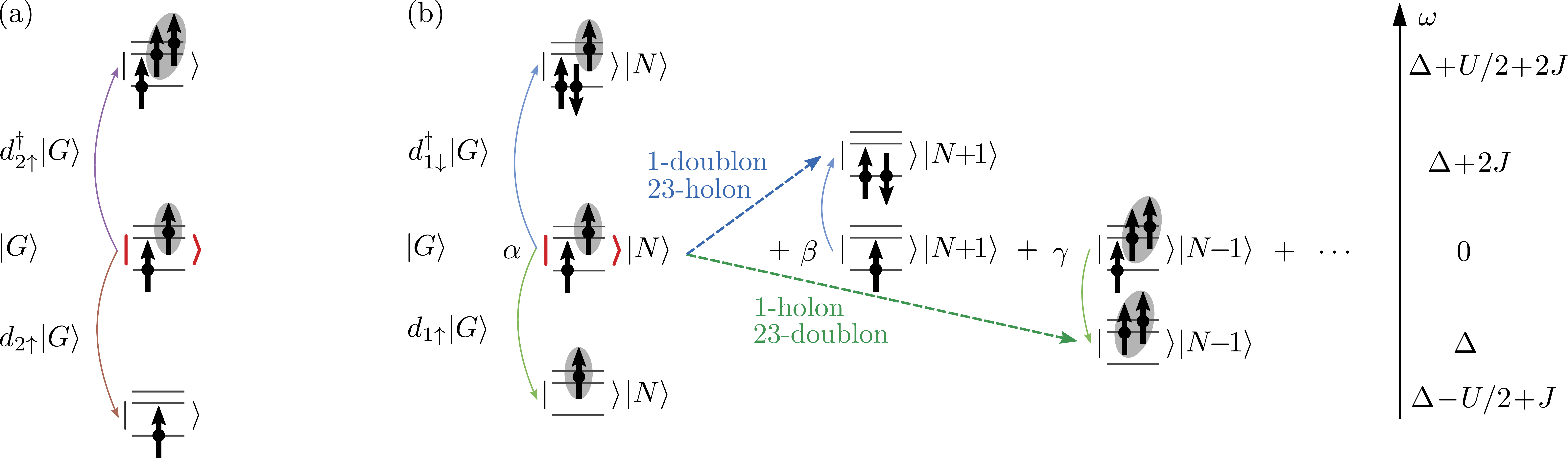}
\caption{%
(a) Ground state $|G\rangle$ in the atomic limit at $\Delta<0$, yet $|\Delta| \lesssim 2J$,
and single-particle and -hole excitations in the $23$-doublet.
Shaded arrows symbolize a symmetric distribution over the degenerate orbitals.
(b) Illustration of interband doublon-holon excitations in the OSMP. 
The occupation of the insulating $1$-orbital is pinned to 1; however, the metallic $23$-doublet still exhibits charge fluctuations.
Then, $|G\rangle$ is a mixture of states, where the dominant impurity occupation is 2
(state marked red), 
and subleading contributions have impurity occupation 1 and 3 ($|\alpha|^2 \gg |\beta|^2, |\gamma|^2$).
At fixed filling, the residual charge is carried by the bath (second ``ket'').
Single-particle and -hole excitations on top of the dominant contribution to the ground state mark the $1$-orbital Hubbard bands. 
Analogous excitations on the subleading terms lead to states which again have impurity occupation 2.
If we relate these states to the dominant first part, 
we can identify them as interband doublon-holon excitations 
\cite{Nunez2018}:
The charge on the impurity remains the same while an electron is removed in the $23$-orbital and added in the $1$-orbital (dashed blue line), or added in the $23$-orbital and removed in the $1$-orbital (dashed green).
The location of the excitations in the $1$-orbital spectral function (right, vertical axis) can be deduced from the atomic energy levels 
[see Eqs.~\eqref{eq:E_HB1} and \eqref{eq:E_dh}].%
}
\label{fig:exc}
\end{figure*}
\section{Doublon-holon excitations}
\label{sec:doublon-holon}
The spectrum of the insulating 1-orbital in the OSMP can be qualitatively explained from the atomic level structure.
In the atomic limit, the ground state consists of eigenstates of the impurity Hamiltonian with one electron in the $1$- and $23$-orbital(s) each 
[the first contribution to $|G\rangle$ in Fig.~\ref{fig:exc}(b), marked in red, is a representative]. 
However, the metallic character of the $23$-orbitals implies charge fluctuations, such that the actual ground state also contains admixtures from states where the $23$-levels of the impurity are empty or doubly occupied [second and third contributions to $|G\rangle$ in Fig.~\ref{fig:exc}(b)].
At fixed filling, the residual charge is carried by the bath [second ``ket'' in the tensor-product notation of Fig.~\ref{fig:exc}(b)].
At large interaction, the first term of $|G\rangle$
with impurity occupation 2 is dominant. 
Single-particle and -hole excitations in the $1$-orbital on top of this state mark the Hubbard bands
[first ``column'' in Fig.~\ref{fig:exc}(b)].
Single-particle and -hole excitations to the other contributions make states accessible which are
inaccessible in the atomic limit 
[second and third ``column'' in Fig.~\ref{fig:exc}(b)]. 
If we relate these states to the \textit{dominant} part of the ground state, 
we can identify them as interband doublon-holon excitations 
\cite{Nunez2018}:
The charge on the impurity remains 2 while an electron is removed in the $23$-orbital and added in the $1$-orbital 
[blue dashed line in Fig.~\ref{fig:exc}(b)] or vice versa (green dashed line).
We can also estimate the positions of both the Hubbard bands
and the doublon-holon peaks in $\mathcal{A}_1$
from the atomic level structure.
To this end, we first recall the impurity Hamiltonian,
$\hat{H}_{\mathrm{imp}}  
= 
\sum_m \epsilon_m \hat{n}_m 
+
\hat{H}_{\mathrm{int}}$,
with
\begin{align}
& H_{\mathrm{int}}  
=
U \sum_m \hat{n}_{m\uparrow} \hat{n}_{m\downarrow}
+
(U\!-\!J)
\sum_{m \neq m'} \hat{n}_{m\uparrow} \hat{n}_{m'\downarrow}
\nonumber \\
& \ 
+
(U\!-\!2J) \sum_{m < m',\sigma} \hat{n}_{m\sigma} \hat{n}_{m'\sigma}
-
J \sum_{m \neq m'} 
\hat{d}^\dag_{m\uparrow} \hat{d}_{m\downarrow}
\hat{d}^\dag_{m'\downarrow} \hat{d}_{m'\uparrow}
\ED
\label{eq:Hint}
\end{align}
The ground-state energy can be estimated from the impurity eigenstate with dominant weight,
having one electron in the $1$-orbital and another spin-aligned one in the $23$-doublet, as
$E_G = \epsilon_1 + \epsilon_{23} + (U-2J)$.
The difference in on-site energies is determined by the crystal field,
$\Delta = \epsilon_1 - \epsilon_{23}$,
and the occupation of $n_1 = 1$ in the OSMP sets a range for their overall shift.
Additionally, a specific value for $\epsilon_{23}$ can be found
by looking at charge fluctuations in the $23$-doublet, as shown next.
\subsection*{Charge fluctuations in the 23-orbitals}
Charge fluctuations in the $23$-doublet on top of the dominant ground-state contribution 
connect the states shown in Fig.~\ref{fig:exc}(a) with atomic energies 
\begin{subequations}%
\begin{align}
E_{\textrm{HB,23}}^+ & = \epsilon_1+2\epsilon_{23} + 3(U-2J),
\\
E_{\textrm{HB,23}}^- & = \epsilon_1
.
\label{eq:dh-1}
\end{align}%
\end{subequations}
The energy cost for the respective transitions,
giving the position of Hubbard bands in the $23$-doublet,
is
\begin{subequations}%
\begin{align}
\delta E_{\textrm{HB},23}^+ & = E_{\textrm{HB},23}^+ - E_G = \epsilon_{23}+2(U-2J), \\
\delta E_{\textrm{HB},23}^- & = E_{\textrm{HB},23}^- - E_G = -\epsilon_{23}-(U-2J).
\end{align}%
\end{subequations}
Equilibrium at filling 2 is thus obtained when 
\begin{equation}
\delta E_{\textrm{HB},23}^+ = \delta E_{\textrm{HB},23}^-
\quad \Rightarrow \quad
\epsilon_{23} = -\tfrac{3}{2}(U-2J).
\label{eq:eps23}
\end{equation}
Inserting the values $U=6$ and $J=1$ mostly used, this means
$\epsilon_{23} = -6$ and $\delta E_{\textrm{HB},23}^\pm = 2$,
corresponding to the bumps in $\mathcal{A}_{23}$
at $\omega = \pm 2$ [Fig.~\ref{fig:sfl_prop}(a)].

\subsection*{Hubbard bands in the 1-orbital}
Single-particle and -hole excitations in the $1$-orbital on top of the dominant ground-state contribution
lead to the states shown in the first ``column'' of Fig.~\ref{fig:exc}(b)
with energies
\begin{subequations}%
\begin{align}
E_{\textrm{HB,1}}^+ & = 2\epsilon_1+\epsilon_{23} + U + (U-J) + (U-2J),
\\
E_{\textrm{HB,1}}^- & = \epsilon_{23}.
\end{align}%
\end{subequations}
Excitations to these states mark the 1-orbital Hubbard bands,
which are found in the spectral function at 
\begin{subequations}%
\begin{align}
\delta E_{\textrm{HB},1}^+ & = E_{\textrm{HB},1}^+ - E_G = \epsilon_1+2U-J, \\
-\delta E_{\textrm{HB},1}^- & = E_G - E_{\textrm{HB},1}^- = \epsilon_1+U-2J.
\end{align}%
\end{subequations}
Inserting the value for $\epsilon_1 = \Delta + \epsilon_{23}$
from Eq.~\eqref{eq:eps23} yields
\begin{subequations}%
\begin{align}
\delta E_{\textrm{HB},1}^+ & = \Delta + \tfrac{1}{2}U + 2J, \\
-\delta E_{\textrm{HB},1}^- & = \Delta - \tfrac{1}{2}U + J.
\end{align}%
\label{eq:E_HB1}%
\end{subequations}
If we further insert the values $\Delta=-1$, $U=6$, and $J=1$
of Fig.~\ref{fig:sfl_prop}(a), we get the peak positions
$-3$ and $4$.
Increasing $U$ up to $8$, with $J=U/6$ as in Fig.~\ref{fig:sfl_prop}(b),
increases their magnitude up to 
$-3\tfrac{2}{3}$ and $5\tfrac{2}{3}$, respectively.
These numbers match the curves in Fig.~\ref{fig:sfl_prop}(a,b) very well.

\subsection*{Doublon-holon subpeaks}
The doublon-holon excitation energies are found from single-particle or -hole excitations on top of the subleading contributions to the ground state with an empty or doubly occupied $23$-doublet
[second and third ``column'' of Fig.~\ref{fig:exc}(b)].
The atomic energies of the excited states are
\begin{subequations}%
\begin{align}
E_{d_1 h_{23}}^+ & = 2\epsilon_1 + U,
\\
E_{h_1 d_{23}}^- & = 2\epsilon_{23} + (U-2J).
\end{align}%
\end{subequations}
The energy difference to the dominant ground-state contribution 
[dashed lines in Fig.\ref{fig:exc}(b)]
gives the position of the subpeaks in the insulating spectral function.
Using $\epsilon_1 - \epsilon_{23} = \Delta$, we have
\begin{subequations}%
\begin{align}
\delta E_{d_1 h_{23}}^+ & = E_{d_1 h_{23}}^+ - E_G  = \Delta + 2J, \\
-\delta E_{h_1 d_{23}}^- & = E_G - E_{h_1 d_{23}}^- = \Delta.
\end{align}%
\label{eq:E_dh}%
\end{subequations}
Interestingly, these peak positions only depend on the difference of the energy levels, $\Delta$,
and on Hund's coupling, $J$.
Inserting the values for Fig.~\ref{fig:sfl_prop}(a) gives $-1$ and $+1$,
and those for Fig.~\ref{fig:sfl_prop}(b) yield $-1$ and $1+U/3$,
in perfect agreement with the plots.

Both the charge fluctuations in the $23$-doublet and the interband doublon-holon excitations are determined by the same subleading contributions to the ground state (such as the terms with coefficients $|\beta|^2$ and $|\gamma|^2$ in Fig.~\ref{fig:exc}).
Hence, the widths of the quasiparticle peak in the $23$-doublet and the subpeaks in the $1$-orbital are closely tied together.
By increasing $E_{\textrm{at}} = U-2J$, one can then decrease both the widths of the $23$-quasiparticle peak and the $1$-subpeaks. On the other hand, by tuning $\Delta$ and $J$ at constant $E_{\textrm{at}}$, one can shift the positions of the $1$-subpeaks, while the weights of the $23$-quasiparticle peak and the $1$-subpeaks remain roughly the same.
\begin{figure*}[h!]
\includegraphics[width=.98\textwidth]{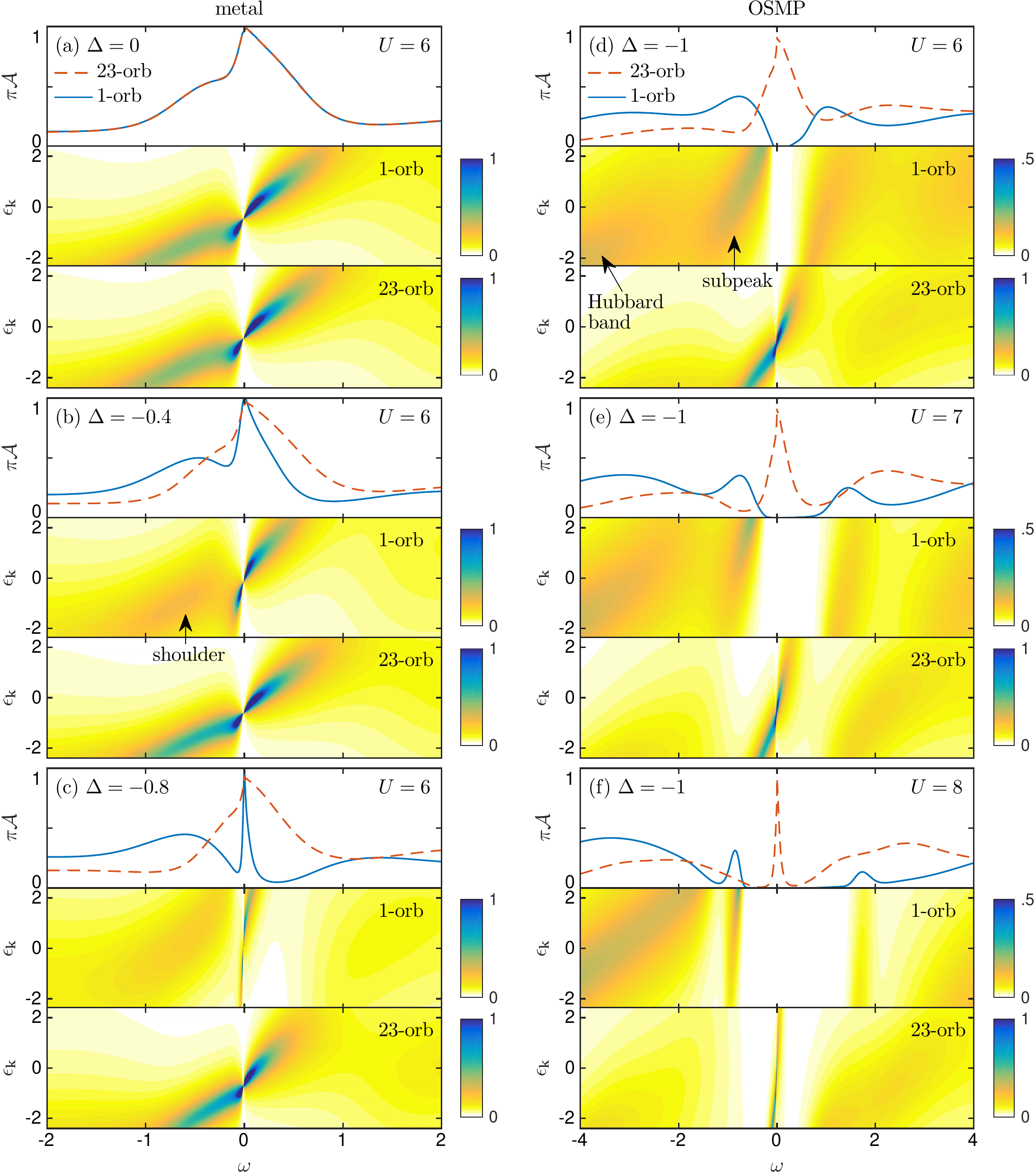}
\caption{%
Local $\mathcal{A}(\omega)$ and momentum-resolved $\mathcal{A}(\omega,\epsilon_{\mathbf{k}})$ spectral functions for varying $\Delta$ in the metallic phase (left panel) and for varying $U$ ($J=U/6$ fixed) in the OSMP (right). Note that, within DMFT, the $\mathbf{k}$ dependence enters only via $\epsilon_{\mathbf{k}}$, and we set the half-bandwidth to 2. (a) Already at $\Delta=0$, $\mathcal{A}(\omega)$ and $\mathcal{A}(\omega,\epsilon_{\mathbf{k}})$ reveal a strong particle-hole asymmetry. (b,c) As we decrease $\Delta$, the 1-orbital is pushed toward half filling, the quasiparticle weight decreases, and $\mathcal{A}(\omega,\epsilon_{\mathbf{k}})$ reveals an almost flat dispersion. Interestingly, the spectral weight from the $\omega<0$ shoulder is continuously transferred from negative to positive $\epsilon_{\mathbf{k}}$. (d) In the OSMP, the quasiparticle weight in the 1-orbital has vanished; the Hubbard band in $\mathcal{A}(\omega,\epsilon_{\mathbf{k}})$ at $\omega<0$ is found at $\epsilon_{\mathbf{k}}<0$ while the subpeak is distinctively centered at $\epsilon_{\mathbf{k}}>0$ (note the altered color scale). The logarithmic singularities in the 23-orbitals are contained in the very sharp structure around $\omega=0$. (e,f) With increasing $E_{\textrm{at}}=U-2J$, the widths of the 23-quasiparticle peak and, consequently, the widths of the 1-subpeaks decrease. With increasing $J$, the position of the right subpeak shifts to higher energies [cf.\ Eq.~\eqref{eq:E_dh}]. The distinct nature of the \textit{interband} doublon-holon excitations and the Hubbard bands becomes clearly visible; they resemble the \textit{intraband} doublon-holon subpeaks in the single-orbital strongly correlated metallic phase 
\cite{Lee2017,Lee2017a}.
}
\label{fig:Akw}
\end{figure*}
\section{Momentum-resolved spectral function}
\label{sec:mom-res-A}
In Fig.~\ref{fig:Akw}, we plot the local spectral function, $\mathcal{A}(\omega)$,
together with the momentum-resolved one, $\mathcal{A}(\omega,\epsilon_{\mathbf{k}})$.
As explained in the caption,
strong particle-hole asymmetry, decreasing quasiparticle weight, and localization of the 1-electrons can be nicely seen.
Moreover, it is interesting to observe that the crossover between the $\omega<0$ shoulder and the interband doublon-holon subpeak at $\Delta$ is accompanied by a transfer of spectral weight from $\epsilon_{\mathbf{k}}<0$ to $\epsilon_{\mathbf{k}}>0$.
In the OSMP, the doublon-holon subpeak at $\omega<0$, $\epsilon_{\mathbf{k}}>0$ can be very well distinguished from the Hubbard band at $\omega<0$, $\epsilon_{\mathbf{k}}<0$.
Especially in the momentum-resolved plot, these \textit{interband} doublon-holon subpeaks resemble the \textit{intraband} doublon-holon subpeaks known from the single-orbital strongly correlated metallic phase 
\cite{Lee2017,Lee2017a}.
\section{Susceptibilities}
\label{sec:susceptibilities}
Here, we give the definitions for the various susceptibilities computed.
The total spin operator is given by
$\hat{\vec{S}} =  \sum_m \hat{\vec{S}}_m$ 
with
$\hat{\vec{S}}_m = \tfrac{1}{2} \sum_{\sigma\sigma'} \hat{d}^{\dag}_{m\sigma} \vec{\tau}_{\sigma\sigma'} \hat{d}_{m\sigma'}$
and Pauli matrices $\vec{\tau}$.
We further define $\hat{\vec{S}}_{23}=(\hat{\vec{S}}_2+\hat{\vec{S}}_3)/2$,
and mainly compute the spin susceptibilities
\begin{equation}
\chi\sp_1=\tfrac{1}{3}\sum_{\alpha=1}^3\langle\hat{S}^\alpha_1 || \hat{S}^\alpha_1\rangle_\omega
, \
\chi\sp_{23}=\tfrac{1}{3}\sum_{\alpha=1}^3\langle\hat{S}^\alpha_{23} || \hat{S}^\alpha_{23}\rangle_\omega
.
\end{equation}%
Further, we use the angular-momentum operator
$\hat{\vec{L}}$ with $\hat{L}_m=i\sum_\sigma\sum_{m'm''} \epsilon_{mm'm''}\hat{d}^{\dag}_{m'\sigma}\hat{d}_{m''\sigma}$
and compute orbital susceptibilities according to
$\hat{L}_{23} = (\hat{L}_2+\hat{L}_3)/2$ and
\begin{equation}
\chi\orb_1=\langle\hat{L}_{23} || \hat{L}_{23}\rangle_\omega
, \quad
\chi\orb_{23}=\langle\hat{L}_1 || \hat{L}_1\rangle_\omega
.
\label{eq:chiorb}
\end{equation}%
In fact, as the system exhibits full SU(2) orbital symmetry in the 23-doublet, we can also use the fully symmetrized 
$\hat{\vec{T}}_{23} = \tfrac{1}{2} \sum_{\sigma}\sum_{m,m'\in\{2,3\}} \hat{d}^{\dag}_{m\sigma} \vec{\tau}_{mm'} \hat{d}_{m'\sigma}$ and
\begin{equation}
\chi\orb_{23}=\tfrac{1}{3}\sum_{\alpha=1}^3\langle\hat{T}^\alpha_{23} || \hat{T}^\alpha_{23}\rangle_\omega
.
\end{equation}%
In the literature, orbital susceptibilities are sometimes computed from charge fluctuations in the individual orbitals. In this language, with $\hat{n}_{23}=(\hat{n}_2+\hat{n}_3)/2$, one has
\begin{equation}
\chi\ch_1 = \langle\hat{n}_1 || \hat{n}_1\rangle_\omega
, \quad
\chi\ch_{23} = \langle\hat{n}_{23} || \hat{n}_{23}\rangle_\omega
.
\end{equation}%
Using orbital SU(2) symmetry with $\chi\orb_{23}=\langle\hat{T}^3_{23} || \hat{T}^3_{23}\rangle_\omega$, one further obtains
\begin{equation}
\tfrac{1}{2}\sum_{m=2}^3\langle\hat{n}_m || \hat{n}_{m}\rangle_\omega
=\langle\hat{n}_{23} || \hat{n}_{23}\rangle_\omega
+\chi\orb_{23}
.
\label{eqn:chi_ch-orb}
\end{equation}%
\begin{figure}[t!]
\includegraphics[width=.37\textwidth]{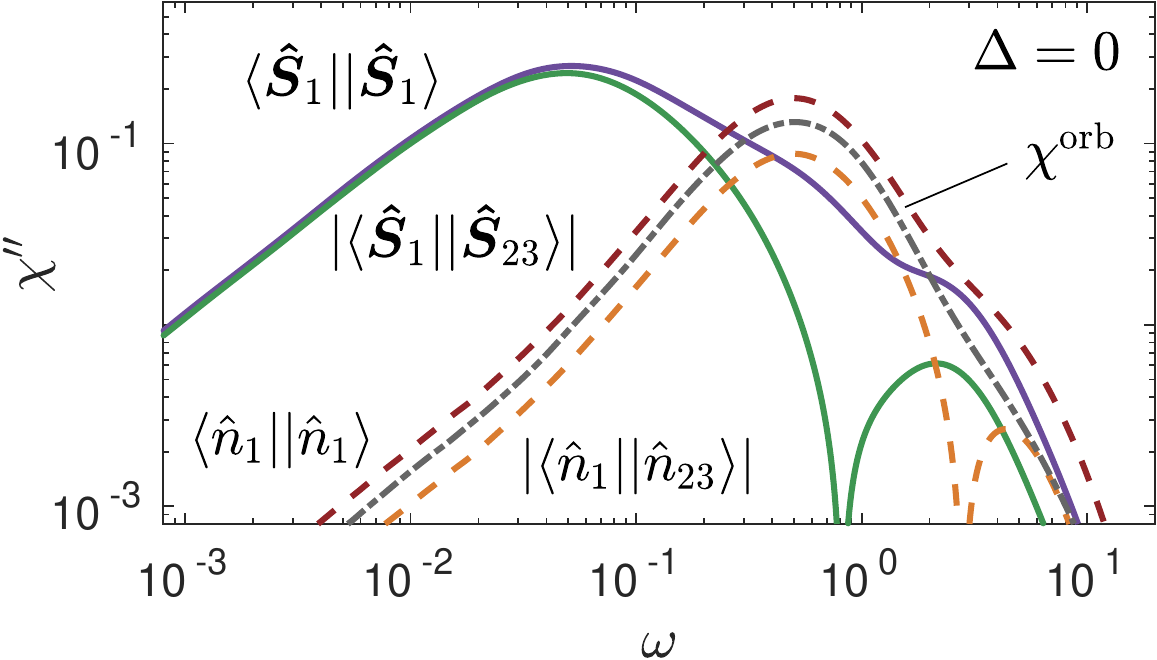}
\caption{%
Various intra- and inter-orbital susceptibilities. As the latter ones change sign within $0<\omega<\infty$, they are shown in absolute value. The orbital Kondo scale can be read off from the position of the maximum of the orbital susceptibility, $\chi\orb$ (dash-dotted line), as well as from orbital-resolved charge susceptibilities, $\langle n_1 || n_1 \rangle_\omega$ and $\langle n_1 || n_{23} \rangle_\omega$ (dashed lines).%
}
\label{fig:app_susceps}
\end{figure}
In the fully symmetric case at $\Delta=0$, we can also extract the spin and orbital 
Kondo temperatures from
\begin{equation}
\chi\sp=\tfrac{1}{3}\sum_{\alpha=1}^3\langle\hat{S}^\alpha || \hat{S}^\alpha\rangle_\omega
, \quad
\chi\orb=\tfrac{1}{8}\sum_{a=1}^8\langle\hat{T}^a || \hat{T}^a\rangle_\omega
,
\end{equation}%
where 
$\hat{\vec{T}} = \tfrac{1}{2} \sum_{\sigma}\sum_{m,m'\in\{1,2,3\}} \hat{d}^{\dag}_{m\sigma} \vec{g}_{m m'} \hat{d}_{m'\sigma}$ with SU(3) Gell-Mann matrices normalized as $\textrm{Tr}[g^a,g^b]=2\delta_{a,b}$.
For illustration, we finally show in Fig.~\ref{fig:app_susceps} intra- and inter-orbital susceptibilities of spin and number operators. As the inter-orbital ones, $\langle \hat{\vec{S}}_1 || \hat{\vec{S}}_{23} \rangle_\omega = \tfrac{1}{3} \sum_{\alpha=1}^3\langle \hat{S}^\alpha_1 || \hat{S}^\alpha_{23} \rangle_\omega$ and $\langle \hat{n}_1 || \hat{n}_{23} \rangle_\omega$, change sign within $0<\omega<\infty$, they are shown in absolute value. 
We see that the orbital Kondo scale, read off from the position of the maximum in $\chi\orb$ (dash-dotted line), can also be determined from orbital-resolved charge susceptibilities (dashed lines),
corresponding to their explicit relation given in Eq.~\eqref{eqn:chi_ch-orb}.
It is interesting to note that spins align, meaning $\langle \hat{\vec{S}}_1 || \hat{\vec{S}}_{23} \rangle_\omega >0$, for $|\omega| \lesssim J = 1$ due to Hund's coupling, and the individual charges antagonize, meaning $\langle \hat{n}_1 || \hat{n}_{23} \rangle_\omega < 0$, for $|\omega| \lesssim U/2 = 3$ to minimize the Coulomb repulsion.
\bibliographystyle{apsrev4-1}
\bibliography{references}

\end{document}